\documentclass[submission,copyright,creativecommons]{eptcs}
\providecommand{\event}{THedu'11} 
\usepackage{breakurl}             
\usepackage{graphicx}

\newtheorem{definition}{Definition}

\title{Automated Generation of User Guidance\\
by Combining Computation and Deduction}
\author{Walther Neuper
\institute{Institute for Software Technology\\
University of Technology
\\
Graz, Austria}
\email{neuper@ist.tugraz.at}
}
\def\titlerunning{Automated Generation of User Guidance}
\def\authorrunning{W. Neuper}
\begin{document}
\maketitle
\begin{abstract}
Herewith, a fairly old concept is published for the first time and named "Lucas Interpretation". This has been implemented in a prototype, which has been proved useful in educational practice and has gained academic relevance with an emerging generation of educational mathematics assistants (EMA) based on Computer Theorem Proving (CTP).
 
Automated Theorem Proving (ATP), i.e. deduction, is the most reliable technology used to check user input. However ATP is inherently weak in automatically generating solutions for arbitrary problems in applied mathematics. This weakness is crucial for EMAs: when ATP checks user input as incorrect and the learner gets stuck then the system should be able to suggest possible next steps.

The key idea of Lucas Interpretation is to compute the steps of a calculation following a program written in a novel CTP-based programming language, i.e. computation provides the next steps. User guidance is generated by combining deduction and computation: the latter is performed by a specific language interpreter, which works like a debugger and hands over control to the learner at breakpoints, i.e. tactics generating the steps of calculation. The interpreter also builds up logical contexts providing ATP with the data required for checking user input, thus combining computation and deduction.

The paper describes the concepts underlying Lucas Interpretation so that open questions can adequately be addressed, and prerequisites for further work are provided. 
\end{abstract}

\section{Introduction}
Motivated by planning for a large project on building educational math assistants (EMAs) at the state-of-the-art of computer mathematics\footnote{MacSchubert http://www.iist.unu.edu/www/docs/annualreports/1993/main\_7.html under founding director Dines Bj{\o}rner}, Peter Lucas\footnote{Peter Lucas was one of the pioneers in compiler construction and in formal methods \cite{pl:formal-lang-hist,pl:form-sem-VDL,pl:lucas70a}. http://www.austria-lexikon.at/af/Wissenssammlungen/Biographien/Lucas\_Peter} supervised the essential design decisions for a ``system that explains itself'', which covers a major part of mathematics as taught to a major target group: solving problems in engineering and in applied sciences in academic courses and at high-schools.

Two requirements were identified that are essential but conflicting: (1) design interaction and notation as close as possible to what is written into textbooks during interactive tutoring and on blackboards during lectures and (2) implement software mechanisms as general and (logically) reliable as possible. The design tackled the conflict in several ways. They key idea is introduced as ``Lucas-Interpretation'' in this paper. Lucas-Interpretation operates on a novel kind of programming language based on Computer Theorem Proving (CTP). A program written in this language determines the next steps while the interpreter builds up logical context providing automated theorem provers (ATP) with facts required to prove user input derivable or not.

The design has been implemented in a prototype called ISAC\footnote{ISAC stands for {\em ISA}belle for {\em C}alculations in Applied Mathematics, http://www.ist.tugraz.at/projects/isac}, first the mathematics engine based on Isabelle~\cite{Nipkow-Paulson-Wenzel:2002}, then a multi-user front-end based on Java-Swing. ISAC has been used successfully in high-schools~ \cite{imst-htl06-SH,imst-htl07-SH,imst-hpts08-SH}. Discussion of underlying concepts has started with focusing education at CADGME~\cite{wn-cadgme09} and continued with focusing technology at THedu'11. The latter workshop confirmed the relevance and (still!) novelty of ISAC's conception. So, after ten years of implementation in ISAC, the underlying concepts herewith are published concisely first time. All concepts described in the paper appear appropriate in ISAC's implementation; {\em not} (yet) implemented features are explicitly designated as such.

\medskip
The paper is structured as follows: \S\ref{running-exp} gives a detailed example which shall motivate the subsequent definitions. \S\ref{deduct} presents how ISAC uses deduction for checking user input, \S\ref{comput} describes how computation is used to construct solutions of problems and \S\ref{ded-comp} introduces Lucas-Interpretation combining deduction and computation for automatic generation of user guidance. Open questions and related work are given ample space in \S\ref{open-related}, and finally \S\ref{summ-concl} gives a summary of Lucas-Interpretation and a conclusion for pedagogy.

\section{Running Example}\label{running-exp}
In order to illustrate the design principles, {\em one} example will be used throughout the paper. This example is from a problem-class with strong traditions in German speaking countries called 'Extremwert Aufgaben'. The example's complexity is at an intermediate level between high school and university, thus indicating the concepts' usability from early introduction of variables up to mathematics applied in studies of science and of engineering.

\begin{figure} [tb]
\centerline{\includegraphics[width=5.0cm]{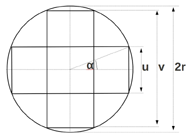}}
\caption{A coil with a cross-shaped kernel}
\label{fig.coil-kernel-uv}
\end{figure}

\paragraph{Example problem:} {\em Given a circle with radius~$r$, where two rectangles with length~$u$ and width~$v$ each are inscribed as shown in Fig.\ref{fig.coil-kernel-uv}. 
This figure shows the section through a coil; the induction current is proportional to the area of the cross-shaped kernel of the coil. Determine $u$ and $v$ such that the kernel's area~$A$ is a maximum.}

\medskip
Traditionally the problem-class 'Extremwert Aufgaben' is taught in calculus courses, thus specification takes geometric considerations 'as obvious' and immediately comes to algebraic formulas as used in the problem's formal specification below. The running example has at least two different specifications:
{\small\begin{tabbing}
123,\=postcond \=: \= $\forall \,A^\prime\, u^\prime \,v^\prime.\,$\=\kill\label{opt-pbl3-spec1}
Specification no.1:\\
\>input    \>: $\{\;r\;\}$  \\
\>precond  \>: $0 < r$   \\
\>output   \>: $\{\;{\it max}\;A,\; u,v\;\}$ \\
\>postcond \>:{\small  $\;A=2uv-u^2 \;\land\; (\frac{u}{2})^2+(\frac{v}{2})^2=r^2 \;\land$}\\
\>     \>\>{\small $\;\forall \;A^\prime\; u^\prime \;v^\prime.\;(A^\prime=2u^\prime v^\prime-(u^\prime)^2 \land
(\frac{u^\prime}{2})^2+(\frac{v^\prime}{2})^2=r^2) \Rightarrow A^\prime \leq A$} \\
\>props\>: $\{\;A=2uv-u^2,\;(\frac{u}{2})^2+(\frac{v}{2})^2=r^2\;\}$
\end{tabbing}
}

Field tests showed, that learners understand what to input in the fields 'input', 'output' and 'props'; the proper postcondition in field 'postcond', however, is out of scope for most students of the target group. The first two fields, 'input' and 'output', are standard, the last one is additional: the field 'props' contains elements of 'postcond', and although the ``logical part'' ($\forall,\exists.\Rightarrow,\land$) is dropped, 'props' distinguishes problems nicely; in fact, the second specification of the running example differs characteristically in 'props', not only in 'postcond':
{\small\begin{tabbing}
123,\=postcond \=: \= $\forall \,A^\prime\, u^\prime \,v^\prime.\,$\=\kill\label{opt-pbl3-spec2}
Specification no.2:\\
\>input    \>: $\{\;r\;\}$  \\
\>precond  \>: $0 < r$   \\
\>output   \>: $\{\;{\it max}\;A,\; u,v\;\}$ \\
\>postcond \>:{\small  $\;A=2uv-u^2 \;\land\; \exists\alpha.\;(\frac{u}{2}=r\sin\alpha \;\land\; \frac{v}{2}=r\cos\alpha) \;\land$}        \\
\>     \>\>{\small $\,\forall \,A^\prime\, u^\prime \,v^\prime.\,(A^\prime=2u^\prime v^\prime-(u^\prime)^2 \,\land\,\exists\alpha^\prime.\;(\frac{u^\prime}{2}=r\sin\alpha^\prime \,\land\,\frac{v^\prime}{2}=r\cos\alpha^\prime)) \Rightarrow A^\prime \leq A$} \\
\>props\>: $\{\;A=2uv-u^2,\;\frac{u}{2}=r\sin\alpha,\;\frac{v}{2}=r\cos\alpha\;\}$
\end{tabbing}
}
In these two examples the fields 'postcond' have the same structure; without having this implemented in ISAC{} yet, we expect that postcondition can be generated automatically from the equations in 'props' and a pattern of the postcondition's structure for the respective problem class. The following definition is (almost) standard.
\begin{definition}[Specification]\label{def-spec}
A quintuple $(I,p(I),O,q(I,O),r(I,O,A))$ is called a specification where $I$ is a set of variables (the \textbf{input variables}), $p(I)$ is a formula in which only the variables from $I$ occur freely (the \textbf{precondition}), $O$ is a set of variables (the \textbf{output variables}) such that $I\cap O = \{\}$, $q(I,O)$ is a formula in which only the variables from $I \cup O$ occur freely (the \textbf{postcondition}) and $r(I,O,A)$ is a set of elements of $q(I,O)$ without quantifiers such that
$$\forall I,O.\;p(I)\Rightarrow(q(I,O)\Rightarrow \exists A.r(I,O,A))$$
is called \textbf{properties}. A vector of terms $S(I)$ in which only $I$ occurs freely (the \textbf{solutions}) solves the specified problem, if the following holds:
$$\forall I.\; p(I) \Rightarrow\forall O.\;O=S(I)\Rightarrow q(I, O)$$
\end{definition}
The specification's elements $I, p(I), O, q(I, O), r(I,O,A)$ are labeled by 'input', 'precond', 'output', \\'props' respectively in the two above example specifications. $A$ in 'props' contains the variables which were existentially bound in $q(I,O)$. 

\bigskip
Given a specification as an {\em implicit} definition of output values, a \textbf{calculation} is a transformation to an {\em explicit} definition by stepwise construction of output values. For instance, give specification no.2 from above, a calculation might be given by the steps in lines {\rm \#01..\#20} below. The calculation as shown below is the result of various experiments trying to accomplish the two conflicting requirements mentioned in the introduction, (1) traditional notation and (2) reliable mechanized treatment.

The details of the calculation below created by ISAC{} will be explained later together with the mechanisms which create the steps and which check the steps if input by the learner. Reactions of students and teachers to this format suggest to just explain what lists like \textit{[maximum\_by, calculus]} or \textit{[make, diffable, function]} mean: In ISAC{} the question for the lists is answered interactively by following a link exactly to the respective specification. Here on paper a link to all specifications in ISAC{} is given\footnote{http://www.ist.tugraz.at/projects/isac/www/kbase/pbl/index\_pbl.html}.
{\small\it\begin{tabbing}
123l\=123\=123\=123\=123\=123\=123\=123\=123\=123\=123\=123\=\kill\label{exp-calc}
\>{\rm 01}\> $\bullet$\> {\tt Problem } [maximum\_by, calculus]       \`- - -\\
\>{\rm 02}\>\> $\vdash$\> $A = 2\cdot u\cdot v - u^2$                 \`- - -\\
\>{\rm 03}\>\> $\bullet$\> {\tt Problem } [make, diffable, function]  \`- - -\\
\>{\rm 04}\>\> \dots\> $\widetilde{A}(\alpha) = 8\cdot r^2\cdot\sin\alpha\cdot\cos\alpha - 4\cdot r^2\cdot(\sin\alpha)^2$  \`- - -\\
\>{\rm 05}\>\> $\bullet$\> {\tt Problem } [on\_interval, max, argument]\`- - -\\
\>{\rm 06}\>\>\> $\vdash$\> $\widetilde{A}(\alpha) = 8\cdot r^2\cdot\sin\alpha\cdot\cos\alpha - 4\cdot r^2\cdot(\sin\alpha)^2$  \\
       \`{\tt Subproblem} [differentiate, function]\\
\>{\rm 07}\>\>\> $\bullet$\> {\tt Problem } [differentiate, function]\\
       \`{\tt Apply\_Method} Differentiate\\
       \`{\tt Take} $\frac{d}{d\alpha}( 8\cdot r^2\cdot\sin\alpha\cdot\cos\alpha - 4\cdot r^2\cdot(\sin\alpha)^2)$\\
\>{\rm 08}\>\>\>\> $\vdash$\> $\frac{d}{d\alpha}( 8\cdot r^2\cdot\sin\alpha\cdot\cos\alpha - 4\cdot r^2\cdot(\sin\alpha)^2)$\\
       \`{\tt Rewrite} $\frac{d}{dx}(a\cdot u-b\cdot v)=a\cdot\frac{d}{dx}u -b\cdot\frac{d}{dx}v$\\
\>{\rm 09}\>\>\>\> $\equiv$\> $8\cdot r^2\cdot\frac{d}{d\alpha}(\sin\alpha\cdot\cos\alpha) - 4\cdot r^2\cdot\frac{d}{d\alpha}(\sin\alpha)^2$\\
       \`{\tt Rewrite} $\frac{d}{dx}(u\cdot v) = u\cdot\frac{d}{dx}v + \frac{d}{dx}u\cdot v$\\
\>{\rm 10}\>\>\>\> $\equiv$\> $8\cdot r^2\cdot(\sin\alpha\cdot\frac{d}{d\alpha}\cos\alpha + (\frac{d}{d\alpha}\sin\alpha)\cdot\cos\alpha) - 4\cdot r^2\cdot\frac{d}{d\alpha}(\sin\alpha)^2$\\
       \`{\tt Rewrite} $\frac{d}{dx}(u^n)=n\cdot u^{n-1}\frac{d}{dx}u$\\
\>{\rm 11}\>\>\>\> $\equiv$\> $8\cdot r^2\cdot(\sin\alpha\cdot\frac{d}{d\alpha}\cos\alpha + (\frac{d}{d\alpha}\sin\alpha)\cdot\cos\alpha) - 4\cdot r^2\cdot 2\cdot(\sin\alpha)^{2-1}\frac{d}{d\alpha}\sin\alpha$\\
\>{\rm 12}\>\>\>\> $\equiv$\> \vdots\\
\>{\rm 13}\>\>\>\> $\equiv$\> $8\cdot r^2\cdot(-(\sin\alpha)^2+(\cos\alpha)^2 - 2\cdot\sin\alpha\cdot\cos\alpha)$\\
       \`{\tt Check\_Postcond} [differentiate, function]\\
\>{\rm 14}\>\>\> \dots\> $\widetilde{A}^\prime(\alpha) = 8\cdot r^2\cdot(-(\sin\alpha)^2)+(\cos\alpha)^2 - 2\cdot\sin\alpha\cdot\cos\alpha)$   \`- - -\\
\>{\rm 15}\>\>\> $\bullet$\> {\tt Problem } [on\_interval, goniometric, equation]\`- - -\\
\>{\rm 16}\>\>\> \dots\> $\widehat{\alpha} = \tan^{-1}(-1+\sqrt{2})$\\
       \`{\tt Check\_Postcond} [on\_interval, max, argument]\\
\>{\rm 17}\>\> \dots\> $\widehat{\alpha} = \tan^{-1}(-1+\sqrt{2})$   \`- - -\\
\>{\rm 18}\>\> $\bullet$\> {\tt Problem} [find\_values, tool]       \`- - -\\
\>{\rm 19}\>\> \dots\> [ $u=0.23\cdot r, \:v=0.76\cdot r$ ]         \`- - -\\
\>{\rm 20}\> \dots\> [ $u=0.23\cdot r, \:v=0.76\cdot r$ ] 
\end{tabbing}}
The numbers on the left margin above do not belong to the calculation, they are for reference in this paper. The strengths of the above format come to bear on interactive worksheets, for instance, the $\bullet$ paired with $\dots$ (on the same level of indentation) indicate places where intermediate steps are collapsed (as in lines {\rm \#03..\#04}) or can be unfolded (as in lines {\rm \#07..\#14}) --- collapsed or unfolded on request by the learner, whether requiring a survey (collapsing and getting the big picture) or whether requiring details (unfolding and pushing parts out of the screen which are not relevant at the moment).

Calculations show two kinds of elements, formulas and tactics. \textbf{Formulas} are shifted to the left margin, which are indented according to a tree structure. \textbf{Tactics} are shifted to the right margin. In the above calculation the tactics are {\tt Subproblem}, {\tt Apply\_Method}, {\tt Take}, {\tt Rewrite} and {\tt Check\_Postcond}; the others are not shown (- - -), according to the tradition to omit on blackboards whatever is not relevant at the moment. 

\medskip
So, the above calculation for the running example provides a first impression, which will motivate rigorous treatment and formal definitions in the sequel.

\section{Deduction for Checking User Input}\label{deduct}
The present prototype checks formulas input by proving equivalence modulo a specified algebraic theory, that means by rewriting to normal forms and checking respective equality. For instance, acceptable inputs at line {\rm \#12} in the above calculation would be those at {\it {\rm \#12} (a...z)} below:
{\small\it\begin{tabbing}
123l\={\rm 12} (b) x\=123\=123\=123\=123\=123\=123\=123\=123\=123\=123\=123\=\kill
\>{\rm 11}\>\> $8\cdot r^2\cdot(\sin\alpha\cdot\frac{d}{d\alpha}\cos\alpha + (\frac{d}{d\alpha}\sin\alpha)\cdot\cos\alpha) - 4\cdot r^2\cdot 2\cdot(\sin\alpha)^{2-1}\cdot\frac{d}{d\alpha}\sin\alpha$\\

\>{\rm 12} (a)\>$\equiv$\> $8\cdot r^2\cdot(\sin\alpha\cdot(-\sin\alpha) + \cos\alpha\cdot\cos\alpha) - 4\cdot r^2\cdot 2\cdot(\sin\alpha)^{2-1}\cdot\frac{d}{d\alpha}\sin\alpha$\\

\>or (b)\>$\equiv$\> $8\cdot r^2\cdot(\sin\alpha\cdot\frac{d}{d\alpha}\cos\alpha + (\frac{d}{d\alpha}\sin\alpha)\cdot\cos\alpha) - 4\cdot r^2\cdot 2\cdot(\sin\alpha)^1\cdot\cos\alpha$\\

\>or (..)\>$\equiv$\> \dots\\

\>or (z)\>$\equiv$\> $8\cdot r^2\cdot(-(\sin\alpha)^2)+(\cos\alpha)^2 - 2\cdot\sin\alpha\cdot\cos\alpha)$
\end{tabbing}}
Each of the inputs {\it(a...z)} has {\it(z)} as a normal form\footnote{Algebraic equivalence modulo {\it thy} of these variants of $f$ is not decidable in general, since there is no normal form for terms containing trigonometric functions \cite{Buchb_Loos:82}; however, in the above case and in most other cases the class of terms is restricted such that ATP reliably can reject or accept an input.}, so each of the inputs is equivalent modulo the simplifier. Although relying on Isabelle \cite{Nipkow-Paulson-Wenzel:2002} as much as possible, ISAC initially implemented a rewrite engine of it's own in order to meet the requirement of transparent single-stepping discussed in \S\ref{inter-steps}. Since there are plans to provide Isabelle's simplifier with appropriate features of transparency\footnote{http://isabelle.in.tum.de/gsoc-ideas.html\#simptrace}, ISAC{} has implemented Isabelle's proof contexts \cite{isar-impl} recently in order to re-use Isabelle's simplifiers.

In Isabelle a proof context is created from an Isabelle theory, i.e. initially it contains all the facts from that theory, and then it is extended dynamicall with facts encountered during interactive proof construction. Isabelle's proof language Isar \cite{mw:isar07} is block-structured and proof contexts follow this structure. So proof contexts contain all {\em locally generated and modified} facts for calling Isabelle's ATP tools, one of which is the simplifier.

\medskip
ISAC{} will re-use the mechanism of Isabelle's proof contexts in order to not only use Isabelle's simplifier, but also Isabelle's other ATP tools. So we claim for a judgment like
\begin{equation} F\vdash_x f \label{check-formula}\end{equation}
where $F$ is the sequence of formulas already present in a calculation, $f$ is a newly input formula and $x$ is a context providing all facts required to derive $f$ via an ATP tool ($\vdash$). $F$ and $f$ are the formulas visible in the calculation, $x$ are the invisible facts additionally required for mechanical proof. \S\ref{ctxt} shows, how $x$ is generated step by step for the calculation of the running example on p.\pageref{exp-calc}.

\subsection{Steps in Derivations}\label{inter-steps}
Experience from educational practice indicates a specific requirement on judgments from ATP:  \textit{judgments shall be presented with comprehensible intermediate steps}. This requirement is hard for ATP, often relying on tools like tableau provers, SMT solvers, on sequent calculi etc., which employ abstract representations hardly understood by novices. However, in case ATP involves simplification (which is frequently the case) this requirement can be successfully accomplished by grouping the rules and showing the rewrites of the groups (and postpone further details to further inquiry). The issue has been acknowledged by CTP development\footnote{Every few weeks the Isabelle mailing list receives a request by a novice for intermediate steps. A ``visual tracing facility for the rewriting engine'' is already addressed at http://isabelle.in.tum.de/gsoc-ideas.html. A related feature exists already in ``proof reconstruction'': data returned from ATP are inserted into Isar proofs in a human readable manner.}, and it seems worth the effort for educational reasons:

When novices are in the phase of acquiring confidence in an EMA, then their curiosity should be decisively supported and questions answered; otherwise EMAs reinforce the wide spread impression of mathematics as magic which cannot be understood. So EMAs should be ``transparent'' to learners' inquiries any time and respond with comprehensible intermediate steps.

\medskip
The requirement for steps coincides with the practical requirement to have some means to stepwise construct a calculation; these means are tactics as shown (partially) in the example calculation on p.\pageref{exp-calc}. Since formulas in a calculation do not contain all information required for mechanical treatment, and this information is contained in contexts, tactics need to operate on contexts. So we have two kinds of tactics, (1) those introduced in the example calculation on p.\pageref{exp-calc} and (2) those explicitly operating on contexts; there is a one-to-one correspondence between (1) and (2), so we need not distinguish between them, except in rare cases: then (1) will be called \textbf{external tactics} and (2) will be called \textbf{internal tactics}.\label{ext-int-tac}

Contexts contain the preconditions $p(I)$ of a specification, the type-constraints given by the input variables $I$ and output variables $O$, assumptions generated and used by conditional rewriting etc. All these facts decide whether a tactic is \textbf{applicable} or not.

Applicability need to be defined for each tactic separately. With respect to the running example on p.\pageref{exp-calc}, tactic {\tt Rewrite} applied at {\rm \#08} requires the fact, that $\widetilde{A}(\alpha)$ is differentiable, and also the left-hand side of the theorem $\frac{d}{dx}(a\cdot u-b\cdot v)=a\cdot\frac{d}{dx}u -b\cdot\frac{d}{dx}v$ must match a redex in the respective formula. Tactic \textit{{\tt Subproblem} [tool, find\_values]} is applicable, if the input variables of the respective specification have been generated in the calculation already, and if the precondition holds. There are also more complicated tactics, for instance handling case-distinctions, which do not show up in the running example. The following definition seems straight forward.

\begin{definition}[Step of calculation]\label{def-calc-step}
Given a specification $s=(I,p(I),O,q(I,O),r(I,O,A))$, a sequence $F$ of formulas, a formula $f^\prime$, contexts $x$ and $x^\prime$ and a tactic $t$, then a step is described by the rule
$$
\displaylines{\hfill
{{t\;{\it is\_applicable\_in}\;(x,F) \qquad  
  x\subseteq x^\prime                  \qquad  
  F\vdash_x x^\prime                   \qquad  
  F\vdash_{x^\prime} f^\prime                  } 
\over 
 {x,F \rightarrow^t x^\prime,F\oplus f^\prime  }}\hfill
}$$
where $F\oplus f^\prime$ means insertion of $f^\prime$ into $F$ at a position according to $t$. The rule is applied to the initial configuration $x_0,F_0$ with tactic $t_o$, were $x_0=p(I)\cup x_{{\it types}(I,O)}\cup x_{\it thy}$, i.e. $x_0$ consists of the precondition, the type constraints $x_{{\it types}(I,O)}$ and the context $x_{\it thy}$ initialized by theory {\it thy}, $F_0=\emptyset$ and $t_0$ is either {\tt Take} or {\tt Subproblem}.

Tactics are designed functional such that for all $t,x,F$ there exists exactly one $x^\prime,F^\prime$ such that $x,F\rightarrow^t x^\prime,F^\prime$. So we can say $\mathbf{t}\;\textbf{applied\_to}\;\mathbf{(x,F)}=\mathbf{(x^\prime,F^\prime)}$
\end{definition}

This definition does not tell whether a step is triggered by input of formula $f$, by input of a tactic $t$ or something else. Although steps are functional, calculations generated by these steps are non-deterministic, because there may be many tactics applicable at one and the same configuration; this variability of calculations is the challenge for learners when deciding on the next step of calculation. \S\ref{ded-comp} will tell details about input of formulas and/or tactics after some further prerequisites have been introduced.

First follows a definition of calculation. The above approach taken via steps suggests in operational view according to \cite{op-sem}:
\begin{definition}[Calculation]\label{def-calc}
Given a specification $s=(I,p(I),O,q(I,O),r(I,O,A))$, then a calculation is a labeled terminal transition system $c=\langle X,{\cal P}(F),T\rightarrow,S\rangle$, where the set of configurations is $X\times {\cal P}(F)$, $X$ a set of \textbf{contexts} and ${\cal P}(F)$ a set of sequences $F$ of \textbf{formulas}, the actions $t\in T$ are called \textbf{tactics}, the transitions in $\rightarrow^t\;\subseteq(X\times {\cal P}(F))\times T\times(X\times {\cal P}(F))$ are the steps of calculation introduced in Def.\ref{def-calc-step}.

The terminal configurations in set $S\subset X\times {\cal P}(F)$ are called \textbf{solutions} of $c$.
\end{definition}
Further details on {\em solutions} are given in \S\ref{solution}.

\bigskip
Summarizing, {\em a calculation consists of steps each of which is derived from the steps previously constructed}; derivation of the steps relies on logical facts (in contexts $x$) not shown to the learner without request. The internal machinery handling the contexts $x$ will be introduced in the subsequent sections.

\subsection{Checking Solutions of Problems}\label{solution}
This subsection presents the only part of the introduced concepts, which is not yet implemented in ISAC{}. This part raises open questions, but is essential for ``a system which explains itself'': such a system should not only construct a calculation stepwise as described in the previous section, such a system also should show that the result of the calculation is a solution of the specified problem. This means (in analogy to judgment (\ref{check-formula}) on p.\pageref{check-formula}) we want to have a judgment
\begin{equation} F\vdash_x q(I,O) \label{check-result}\end{equation}
where $F$ are the formulas in the (completed) calculation, $x$ is the final context and $q(I,O)$ is the postcondition on the input $I$ and the output $O$ in the respective specification. 

A look at the running example sheds light on the challenges for designing the details for an implementation of (\ref{check-result}). In the example $p(I,O)$ is
{\small\begin{tabbing}
123l\=postcond \=: \= 1234\=\kill
\>postcond \>:{\small  $\;A=2uv-u^2 \;\land\; \exists\alpha.\;(\frac{u}{2}=r\sin\alpha \;\land\; \frac{v}{2}=r\cos\alpha) \;\land$}        \\
\>     \>\>{\small $\,\forall \,A^\prime\, u^\prime \,v^\prime.\,(A^\prime=2u^\prime v^\prime-(u^\prime)^2 \,\land\,\exists\alpha^\prime.\;(\frac{u^\prime}{2}=r\sin\alpha^\prime \,\land\,\frac{v^\prime}{2}=r\cos\alpha^\prime)) \Rightarrow A^\prime \leq A$} 
\end{tabbing}
}
The postcondition's $\exists$ is accomplished by constructing $\alpha$ in the calculation's lines {\rm \#15--\#16}. A complete proof would have to re-curse to the postcondition of (Sub-)\textit{{\tt Problem} [on\_interval, max, argument]} at line {\rm \#05} which might look like 
$$\forall \alpha.\; \alpha\in\;]0,\frac{\pi}{2}[ \;\Rightarrow \widetilde{A}(\alpha) \leq \widetilde{A}(\tan^{-1}(-1+\sqrt{2}))$$
A closer look suggests to have not only this postcondition in the context $x$, but also the function $\widetilde{A}(\alpha)$, the facts $A = \widetilde{A}(\tan^{-1}(-1+\sqrt{2}))$, $u=0.23$, $v=0.76$ etc. --- so (automatically!) proving the final postcondition needs to rely on a well-stocked context $x$; building $x$ is the main concern of \S\ref{comput}.

\medskip
Since we have introduced a calculation as ``consisting of steps each of which is derived from the previous steps'' the steps need to be considered. The final step in the example calculation is in line {\rm \#19}, obtaining the values \textit{[$u=0.23\cdot r, \:v=0.76\cdot r$]} as results of (Sub-)\textit{{\tt Problem} [tool, find\_values]}. These values are approximations of exact terms which are too large to be carried through a calculation by hand; due to the design decision for notation ``as close to what is written to traditional blackboards as possible'' the running example also uses approximate values --- however, approximate values cannot prove the postcondition given above; the postcondition would require re-formulation by using the notion of ``approximation errors'', which would make the postcondition and the proof even more complicated.

\medskip
An implementation for checking solutions shall be analogous to Def.\ref{def-calc-step} on stepwise construction of calculations, i.e. there should be a final step which completes the calculation such that the postcondition holds:
\begin{definition}[Result of calculation]\label{def-result}
Given a specification $s=(I,p(I),O,q(I,O),r(I,O,A))$ and a calculation $c=\langle X,{\cal P}(F),T,\rightarrow,S\rangle$ as a labeled terminal transition system with transitions according to Def.\ref{def-calc-step} for steps of calculation with $x_0,\emptyset\rightarrow^{t_0}\dots\rightarrow^{t_n} x_n,F_n\rightarrow^{\tt Check\_Postcond} x_n,F_n$ such that
$$F_n\vdash_{x_n}q(I,O)$$
and for every variable $o\in O$ there is an equation $o=r$ in $F_n$ such that all the equations can be ordered in a sequence
$$o_1=r_1, o_2=r_2, \dots, o_n=r_n$$
where every $o_i$ does not occur in $r_1,\dots,r_i$. Then $r_1,\dots,r_n$ is called a \textbf{result} of $c$ and $x_n,F_n$ is called a \textbf{solution} constructed by $c$. We also say that calculation '$\mathbf{c}\;\textbf{is\_completed\_with}\;\mathbf{(x_n,F_n)}$'. 
\end{definition}

\section{Computation for Constructing Solutions}\label{comput}
Since computers exist, they are used to compute some output from some input, where the resulting output fulfills certain expectations --- i.e. they are used just for the purpose formalized in the above Def.\ref{def-result} for solutions of computations. Symbolic computation allows to handle formal objects of arbitrary level of abstraction, even real numbers like $\tan^{-1}(-1+\sqrt{2})$ in {\em finite} computers ($\sqrt{2}$ etc. not as a floating point number, of course).

\begin{figure} [tb]
\centerline{\includegraphics[width=8.0cm]{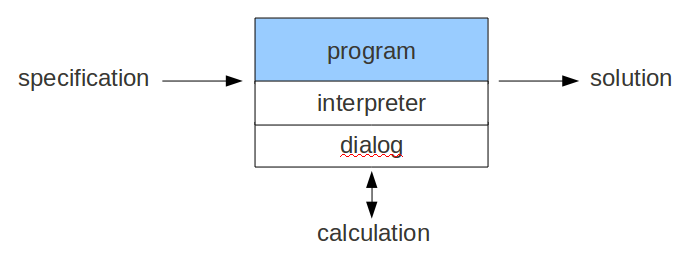}}
\caption{A calculation is a side-effect of a functional program}
\label{fig.lucin-sideeffect}
\end{figure}

\medskip
Computation envisaged in this section is used to create steps of calculations. ISAC{}'s design provides a functional programming language for doing that computation; given the characteristic of functional programs to produce a result (without intermediate states), the question arises, how intermediate steps can be created such that interaction on these steps is possible (since this clearly requires states related to these steps). Fig.\ref{fig.lucin-sideeffect} on p.\pageref{fig.lucin-sideeffect} 
gives a preliminary answer to this question: the calculation and respective steps are created as a side-effect of interpreting the program; the states and interaction on the steps of calculation are handled by the interpreter of the program. The role of a dialog in such an architecture seems natural, but details are out of scope in this paper.

\subsection{CAS-like Functionality in Programming Languages}\label{cas-like-funct}
For constructing calculations we need functionality well-known from Computer Algebra Systems (CAS): for instance, differentiation and equation solving in the running example on p.\pageref{exp-calc}. Indeed, CAS-based programming languages \cite{prog-maple06,progr-mathematica} are most successful in software production for engineering and science. However, CAS-based languages do not commend themselves for educational math assistants --- users of CAS are responsible for when to apply CAS functions and for how to use their results; so learners might be over-challenged by such responsibility: equation solvers might have lost some solutions, simplification might imply partiality conditions without mentioning them (e.g. $\frac{x^2-1}{x-1}=x+1$ without mentioning $x\not=1$) etc. For education we want to have systems ``which never make mistakes'' \cite{cezary-phd}.

Deduction is already in the game since \S\ref{deduct}, so we advocate rigor corresponding to CTP: we claim any CAS-function to be accompanied by an explicit formal specification. Coming back to the running example, differentiation as shown for the example calculation in lines {\rm \#07..\#13} on p.\pageref{exp-calc} is specified by
{\small\begin{tabbing}
123\=123,\=postcond \=: \= \=\kill\label{spec-deriv}
\>Specification of {\it{\tt Problem } [differentiate, function]}:\\
\>\>input    \>: \>$\{\;f,v\;\}$  \\
\>\>precond  \>: \>$\lambda v.\:f\;{\it is\_differentiable}$   \\
\>\>output   \>: \>$\{\;f^\prime, {\it range}\;\}$ \\
\>\>postcond \>: \>$\forall v.\;v\in{\it range}\;\Rightarrow (\lambda v.\:f^\prime)(v)=\lim_{h\rightarrow 0}\frac{(\lambda v.\:f)(v+h)-(\lambda v.\:f)(v)}{h}$
\end{tabbing}\label{spec-diff}
}
\noindent
and this implicit specification is transformed to an explicit specification, i.e. a calculation, by the tactic {\it{\tt Subproblem} [differentiate, function]}: in line {\rm \#07} this tactic starts the Sub-{\it{\tt Problem }[differentiate, function]} and solves the problem by tactic {\it{\tt Apply\_Method} Differentiate}. The respective program is shown on p.\pageref{prog-diff} below.

\medskip
The above requirements advocate a novel kind of programming called ``CTP-based''. Considerable work is already done preparing such languages \cite{cezary-phd,caspartial,casproto}; this is discussed under related work in \S\ref{ctp-based-lang}. 

\subsection{A CTP-based Prototype Language}\label{ctp-based-prog}
ISAC{} implements a programming language, which is appropriate for describing mathematics found in problems of engineering and of science. The following program generates the calculation on p.\pageref{exp-calc} for the running example; at a first glance the reader sees a purely functional program with the usual keywords {\tt LET}..{\tt IN}, {\tt IF}..{\tt THEN}..{\tt ELSE}; at this point the only difference are the function calls: identified by {\tt Subproblem}, these calls are associated with a theory (e.g. \textit{Reals}), a reference to a specification (e.g. \textit{[make, diffable, function]} and a reference to the function itself (e.g. \textit{make\_fun}). 

In the same way the {\tt Subproblem}s are accompanied by theory, specification and program, the example program below is accompanied by a theory and a specification already introduced in two variants (for the same program!) no.1 or no.2 on p.\pageref{opt-pbl3-spec2}. That means, the arguments of {\it{\tt Program} Max} are checked by the respective precondition and the return-value \textit{finds} is related to the input by the respective postcondition on termination of the program:
{\small\it\begin{tabbing}
123l\=123\=123\=123\=123\=123\=123\=123\=123\=123\=123\=123\=123\=\kill\label{prog-max}
\>{\rm 01}\>  {\tt Program} Max (givens::real list) (max::real) (finds::real list) (rels::bool list)\\
\>{\rm 02}\>\>\>\>\>  (var::real) (interval::real set) (errbound::real) =   \\
\>{\rm 03}\>\>  {\tt LET}                                                      \\
\>{\rm 04}\>\>\>  maxequ = {\tt Take} (({\tt HD} o ({\tt FILTER} ({\tt contains} max ))) rels) ;   \\
\>{\rm 05}\>\>\>  funterm = ({\tt IF} $1<$ {\tt LEN} rels                            \\
\>{\rm 06}\>\>\>\> {\tt THEN} ({\tt Subproblem} ( Reals, [make, diffable, function], make\_fun)   \\
\>{\rm 07}\>\>\>\>\>\>\>  [ max::real, var::real, rels::bool list, interval::real set ] )\\
\>{\rm 08}\>\>\>\>  {\tt ELSE} ({\tt HD} rels )) ;                                    \\
\>{\rm 09}\>\>\>  max = {\tt Subproblem} ( Real\_Algebra, [on\_interval, max, argument],\\
\>{\rm 10}\>\>\>\>\>\>\>\>\>\>\>\>              maximum\_on\_interval )\\
\>{\rm 11}\>\>\>\>\>\>\> [ funterm::real, var::real, interval::real set ] ;\\
\>{\rm 12}\>\>\>  find\_rels = {\tt FILTER\_OUT} ({\tt ident} maxequ) rels ;\\
\>{\rm 13}\>\>\>  finds = {\tt Subproblem} ( Reals, [tool, find\_values], find\_values)   \\
\>{\rm 14}\>\>\>\>      [ max::real, ({\tt RHS} funterm)::real, var::real, max::real,\\
\>{\rm 15}\>\>\>\> $\;\;$find\_rels::bool list, errbound::real ] \\
\>{\rm 16}\>\>   {\tt IN } finds
\end{tabbing}}
In ISAC{} this program is parsed as an Isabelle term with {\tt LET}..{\tt IN}, {\tt IF}..{\tt THEN}..{\tt ELSE} from Isabelle/HOL extended by definitions for {\tt Program}, for ``tactics'' like {\tt Take}, {\tt Subproblem} and for some functions {\tt LEN}gth (of a list), {\tt RHS} (right-hand side of an equality), {\tt FILTER\_OUT}; some of these functions are already contained in the recent Isabelle versions.
The reader may note that the notion of ``tactic'' is being extended: Tactics have been introduced in Def.\ref{def-calc-step} as parts of steps in calculations, in the above program certain \textit{statements} are called ``tactic'' --- those statements which generate the steps of the respective calculation. \S\ref{steps-interp} will unify the notions of tactics formally.

The function's arguments are worth one more remark: For instance, the {\tt Subproblem} at lines {\rm \#09}..{\rm \#11} above takes two arguments, a triple and a list; the triple addresses three kinds of knowledge as mentioned above:
\begin{enumerate}
\item deductive knowledge: {\it Real\_Algebra} is an \textbf{Isabelle theory} comprising all the axioms, definitions and theorems required for the formulas of the generated calculation and for proving the postcondition of the calculation's specification. This is where the differentiation rules come from in the second example program on p.\pageref{prog-diff}.
\item application-oriented knowledge: {\it [on\_interval, max, argument]} is a path into a hierarchy of \textbf{specification}s: the element retrieved from the hierarchy specifies the calculation to be generated.

The specification is expected to be proved by ATP during interpretation: before the call the precondition is proved for the function's arguments, on return the postcondition is proved for arguments and return value.
\item  algorithmic knowledge: {\it maximum\_on\_interval} is a reference to the \textbf{program} describing the algorithm which creates the calculation.\\
\end{enumerate}

The second argument of the {\tt Subproblem} contains some formal arguments {\it funterm, var, interval} as known from function definitions: 
{\small\it\begin{tabbing}
123\=123\=123\=123\=(\=123\=123\=123\=123\=123\=\kill
\>{\it [ funterm::real, var::real, interval::real set ]}
\end{tabbing}}\label{prog-diff}
The list of arguments might contain more items than the respective specification --- this becomes clear when considering the example program above which has these formal arguments: 
{\small\it\begin{tabbing}
123l\=123\=123\=123\=123\=123\=123\=123\=123\=123\=123\=123\=\kill
\>{\rm 01}\>  {\tt Program} Max (givens::real list) (max::real) (finds::real list) (rels::bool list)\\
\>{\rm 02}\>\>\>\>\>  (var::real) (interval::real set) (errbound::real) = \dots
\end{tabbing}}
The number of arguments is seven, although both respective specifications no.1 and no.2 on p.\pageref{opt-pbl3-spec2} only have one input, $\{\;r\;\}$ for the first argument {\it givens} --- the other arguments provide the input for all the {\tt Subproblem} of the program!

\medskip
This program is the first of two examples for a ``CTP-based programming language'' (where the second example program is on p.\pageref{prog-diff} below). A thorough introduction of such a language is out of scope in this paper, \cite{plmms10} gives further details and \S\ref{ctp-based-lang} discusses open questions.

\medskip
The reader may note, that the above programs are purely functional, thus do {\em neither} they contain input statements {\em nor} output statements, and thus cannot be concerned with interaction\footnote{The programmer of a CTP-based program is much concerned with theories, specifications and algorithms; so excluding concerns of interaction, of user guidance, etc. is a helpful separation of concerns.}. Interaction, however, is the key point of this paper and will be tackled subsequently.

\section{Deduction {\em and} Computation for User Guidance}\label{ded-comp}
Now all prerequisites are prepared for introducing ``Lucas-Interpretation'': specifications of problems, calculations stepwise creating an explicit specification by applying tactics, calculations augmented by logical contexts, contexts containing facts required to check user input and to check the postcondition by ATP, a CTP-based programming language involving theories, specifications and programs, CAS-functionality included in this language.

Now the idea of Lucas-Interpretation is simple:
\begin{itemize}
\item The interpreter works like a debugger jumping from break-point to break-point in a program.
\item The breakpoints are the tactics in the program; each tactics generates exactly one step in the calculation (a step which might comprise an arbitrary number of sub-steps, which are not shown to the user --- rather, the system is ready to show them on request of the user).
\item Each step also generates specific logical facts; the interpreter builds up contexts with these facts according to certain rules.
\item All steps are presented to the user on a ``worksheet'' (if not a dialog component decides for some other kind of interaction, for instance in a written exam).
\item After a step has been presented to the user, control is handed over to the user (to the dialog component, respectively).
  \item Since the user is in control of the system, she or he has a variety of choices collected under four categories:
  \begin{itemize}
  \item Inspect the calculation generated so far: ask for theorems applied at certain steps, ask for intermediate steps for instance in {\it{\tt Rewrite\_Set} simplifier}, etc.
  \item Investigate the underlying knowledge in theory, specification or program: inspection starts at the particular knowledge item used in the current step. 
  \item Input an own step independently, a formula or a tactic (or something else according to the dialog component, for instance, a partial formula presented for filling in just some (crucial) part).
  \item If got stuck in the calculation ask the system to do the next step. This feature enables  a dialog component to provide \textbf{user guidance}; dialog details are out of scope of this paper; see \cite{kremp.np:assess}.
  \end{itemize}
\item User input is checked by ATP, which is called by the interpreter using facts collected in the context. Feedback from ATP goes to the user (probably filtered by the dialog component).
\item After input the interpreter resumes execution, probably at another location in the program due to free user input.
\end{itemize}
The subsequent sections describe first how the interpreter jumps from break-point to break-point, then how logical context are built up during interpretation and finally how user guidance is created automatically. 

\subsection{Steps of Interpretation Create Steps of Calculation}\label{steps-interp}
Steps of calculation (Def.\ref{def-calc-step}) comprise tactics as parts which give a denotative name to the step; within programs those statements are called (program) tactics which create a step of the respective calculation. Programs comprise further statements, which do {\em not} create steps; these are called \textbf{non-generating statements}. 

Since {\it{\tt Program} Max} contains only one non-generating statement in line {\rm \#12}, we give another example: a program calculating the derivative of a function and making the specification on p.\pageref{spec-deriv} explicit:
{\small\it\begin{tabbing}
123l\=123\=123\=123\=(\=123\=123\=123\=123\=123\=\kill\label{prog-diff}
\>{\rm 01}\> {\tt Program} Differentiate (interval::real set) (f::real) (v::real) =\\
\>{\rm 02}\>\>{\tt LET} f' = {\tt Take} ($\frac{d}{d\;{\it v}}$ f)\\
\>{\rm 03}\>\>{\tt IN}\>({\tt REPEAT} \\
\>{\rm 04}\>\>        \>\>\>( {\tt Rewrite\_Inst} [(bdv, v)] diff\_sum ) {\tt OR}\\
\>{\rm 05}\>\>        \>\>\>( {\tt Rewrite\_Inst} [(bdv, v)] diff\_product ) {\tt OR}\\
\>{\rm 06}\>\>        \>\>\>( {\tt Rewrite\_Inst} [(bdv, v)] diff\_sin ) {\tt OR}\\
\>{\rm 07}\>\>        \>\>\>( {\tt Rewrite\_Inst} [(bdv, v)] diff\_cos ) {\tt OR}\\
\>{\rm 08}\>\>        \>\>\>( {\tt Rewrite\_Inst} [(bdv, v)] \dots\dots ) {\tt OR}\\
\>{\rm 09}\>\>        \>\>\>( {\tt Rewrite\_Inst} [(bdv, v)] diff\_fraction )) {\tt @@}\\
\>{\rm 10}\>\>        \>\>( {\tt TRY} ({\tt Rewrite\_Set} simplifier ))$\;$ f'
\end{tabbing}}\label{prog-diff}
Again, the syntax of this kind of program is discussed in detail in \cite{plmms10}, here are the hints for reading the program text: After constructing the formula from the program's arguments in line {\rm \#02} the body of {\tt LET..IN} forward chaines ({\tt @@}) two functions, {\tt REPEAT}$(\dots)$ and {\it {\tt TRY} ({\tt Rewrite\_Set} simplifier)}. The former models the non-deterministic behavior of term rewriting systems: the rules {\it diff\_sin {\tt OR} diff\_cos {\tt OR} \dots}\footnote{In order to come traditional notation as close as possible, the type of the function term is \textit{real} and not ${\it real}\Rightarrow{\it real}$; so we need to instantiate {\it bdv} by the actual value of {\it v} in the rules before rewriting.} or other rules can apply or not, and as soon none applies {\tt REPEAT} terminates. The latter function {\tt TRY}s to simplify the resulting formula.

The lines {\rm \#08--\#13} in the example calculation on p.\pageref{exp-calc} could have been generated by this program. Let us assume this calculation has been continued as follows:
{\small\it\begin{tabbing}
1234\=123\=123\=123\=123\=123\=123\=123\=123\=123\=123\=123\=\kill
\>{\rm 11}\> $\equiv$\> $8\cdot r^2\cdot(\sin\alpha\cdot\frac{d}{d\alpha}\cos\alpha + (\frac{d}{d\alpha}\sin\alpha)\cdot\cos\alpha) - 4\cdot r^2\cdot 2\cdot\sin\alpha\cdot\frac{d}{d\alpha}\sin\alpha$\\
\`{\tt Rewrite} $\frac{d}{d\alpha}(\sin{\alpha})=\cos{\alpha}\;\;$ {\rm \#06}\\

\>\> $\equiv$\> $8\cdot r^2\cdot(\sin\alpha\cdot\frac{d}{d\alpha}\cos\alpha + \cos\alpha\cdot\cos\alpha) - 4\cdot r^2\cdot 2\cdot\sin\alpha\cdot\frac{d}{d\alpha}\sin\alpha$\\
\`{\tt Rewrite} $\frac{d}{d\alpha}(\cos{\alpha})=-\sin{\alpha}\;\;$ {\rm \#07}\\

\>\> $\equiv$\> $8\cdot r^2\cdot(\sin\alpha\cdot(-\sin\alpha) + \cos\alpha\cdot\cos\alpha) - 4\cdot r^2\cdot 2\cdot\sin\alpha\cdot\frac{d}{d\alpha}\sin\alpha$\\
\`{\tt Rewrite} $\frac{d}{d\alpha}(\sin{\alpha})=\cos{\alpha}\;\;$ {\rm ???}
\end{tabbing}}
First line {\rm \#06} of the program has been interpreted (indicated by the line number at the right margin above), then the subsequent line {\rm \#07}. Finally line {\rm \#06} with the rule {\it diff\_sin} should be interpreted again in order to derive $\frac{d}{d\alpha}\sin\alpha$ --- how does the interpreter come back to the respective location in the program?

Starting from location {\rm \#07}, this is accomplished by the sequence of statements
{\small\begin{tabbing}
123l\=123\=123\=\kill
\>${\rm \#07}\rightarrow^{\tt OR}\dots\rightarrow^{{\tt Rewrite}\;{\it diff\_fraction}}{\rm \#09}\rightarrow^{\tt REPEAT}{\rm \#03}\rightarrow^{{\tt Rewrite}\;{\it diff\_sum}}{\rm \#04}$\\
\>${\rm \#04}\rightarrow^{\tt OR}{\rm \#05}\rightarrow^{{\tt Rewrite}\;{\it diff\_product}}{\rm \#05}$\\
\>${\rm \#05}\rightarrow^{\tt OR}{\rm \#06}\rightarrow^{{\tt Rewrite}\;{\it diff\_sin}}$
\end{tabbing}
}
The statements {\tt OR}, {\tt REPEAT} (and also {\tt LET..IN}, {\tt @@}, {\tt Try}) are concerned with controlling the sequence of interpretation and are called \textbf{tacticals}; they do {\em not} change the calculation or the respective context. ISAC{}'s interpreter controls such sequences by locations in the program and by environments as usual. 

The above example leads to the following definition in a straight forward manner, just adding a program-state to Def.\ref{def-calc-step} for steps of calculation:
\begin{definition}[Step of interpretation]\label{def-interp-step}
Given a specification $s=(I,p(I),O,\_,\_)$, a program $m$ and a respective program-state $\gamma$, given non-generating statements $s_{\it ng}$ and a (program)tactic $t_m$ both in $m$, given an (internal)tactic $t_c$ according to Def.\ref{def-calc-step}, a context $x$ and a sequence of formulas $F$, then a \textbf{step} is an application of the rule
$$
\displaylines{\hfill
{{\gamma\rightarrow^{s_{\it ng}}\dots\rightarrow^{s_{\it ng}^\prime}\gamma^\prime\rightarrow^{t_m}\gamma^{\prime\prime}
     \qquad t_m\approx t
     \qquad t\;{\it applied\_to}\;(x,F)=(x^\prime,F^\prime) 
} 
\over 
 {\gamma,x,F \rightarrow^{t} \gamma^{\prime\prime},x^\prime,F^\prime  }}\hfill
}$$
where $t_m\approx t_c$ denotes equivalency induced by a bijective mapping between program tactics and tactics in calculations -- due to this equivalency the distinction is omitted in the rule's conclusion and the tactic named $t$. 

The initial configuration for applying the rule is $\gamma_0,x_0,F_0$ with $\gamma_0=(\emptyset,\{(i_1,v_1),\dots,(i_n,v_n)\})$ where $\emptyset$ is the location at the root of program $m$ and $(i_i,v_i)$ are the identifier-value pairs of the arguments of $m$, and $x_0,F_0$ are as defined in Def.\ref{def-calc-step}. We say 'the steps are $\textbf{initialized\_by}\;\mathbf{s,m}$'.
\end{definition}
Interpretation stops at the first applicable tactic found after passing non-generating statements $s_{\it ng}$, which are either tacticals or expressions containing no tactic. A subsequent step is triggered from outside without caring about program tactics, so we introduce a judgement \textbf{do\_next} generalizing over all tactics~$t$
$$
\displaylines{\hfill
{{\gamma,x,F\rightarrow^t \gamma^\prime,x^\prime,F^\prime
} 
\over 
 {\gamma,x,F \rightarrow^{\it do\_next} \gamma^\prime,x^\prime,F^\prime  
}}\hfill
                                               \llap{{\it do\_next}}
}$$

A detailed description of the tacticals' semantics is out of scope of this paper. We just note, that the example program on p.\pageref{prog-diff} would raise an error in line {\rm \#10} if the tactical {\tt TRY} would be omitted and the tactic {\it{\tt Rewrite\_Set} simplifier} would be {\em not} applicable.

\subsection{Building up Logical Context during Interpretation}\label{ctxt}
Contexts are required to check user input (\S\ref{deduct}) by ATP during interpretation, to check preconditions and post-conditions of {\tt Subproblem}s  and finally check the solution of a problem (\S\ref{solution}) by ATP as well. How contexts are used to accomplish these tasks is introduced in \S\ref{lucas-interp}.

How logical contexts $x_i$ are built up during execution can be observed in the running example as follows (the numbers on the left refer to those in program on p.\pageref{prog-max}):
{\small\begin{tabbing}
123l\={\rm 01..02} \=(a)\= $c_0 = \{$\=123\=123\=123\=123\=123\=123\=123\=123\=123\=\kill
\>{\rm 01..02} \>\>$x_0 = \{\; 0 < r \;\}\cup x_{{\it types}(r,A,u,v)}\cup x_{\it thy}$\\

\>{\rm 03..04} \>\> $x_1 = x_0 \cup \{\; A=2uv-u^2 \;\}$\\

\>{\rm 05..08} \>\> $x_2 = x_1 \cup \{\; \widetilde{A}(\alpha) = 8\cdot r^2\cdot\sin\alpha\cdot\cos\alpha - 4\cdot r^2\cdot(\sin\alpha)^2), $\\
\>\>\>\>\> $\;\;\widetilde{A}(\alpha)\;{\it is\_differentiable\_on}\;]0,\frac{\pi}{2}[ \;\;\}$\\

\>{\rm 09..12} \>{\rm a}\> $x_3 = x_2 \cup \{\; \widetilde{A}^\prime(\alpha) = 8\cdot r^2\cdot(-(\sin\alpha)^2)+(\cos\alpha)^2 - 2\cdot\sin\alpha\cdot\cos\alpha),$\\
\>\>{\rm b}\>\>\> $\;\;\alpha = \tan^{-1}(-1+\sqrt{2}),\;
      \widetilde{A}^\prime(\alpha) = 0,\;$\\
\>\>{\rm c}\>\>\> $\;\;\forall\alpha^\prime.\;\alpha^\prime\in\;]0,\frac{\pi}{2}[ \;\land\, \widetilde{A}^\prime(\alpha^\prime)=0 \;\Rightarrow\; \alpha^\prime=(\tan^{-1}(-1+\sqrt{2})) $\\
\>\>{\rm d}\>\>\> $\;\;\forall \alpha^\prime.\; \alpha^\prime\in\;]0,\frac{\pi}{2}[ \;\Rightarrow \widetilde{A}(\alpha^\prime) \leq \widetilde{A}(\alpha) \;\}$\\

\>{\rm 13..15} \>\> $x_4 = x_3 \cup \{ u=2\cdot r\cdot\sin\alpha,\; u\approx 0.23\cdot r,\;
    v=2\cdot r\cdot\cos\alpha,\; u\approx 0.76\cdot r \;\}$

\end{tabbing}
}
The initial values in $x_0$ comprise the precondition and $0 < r$, the type constraints for the input variables $I$ and the output variables $O$ of the specification as well as all required knowledge collected in theory {\it thy}.

The facts added in lines {\rm \#09..\#12} to context $x_2$ deserves explanation: the first line (a) is the result of Sub-{\it{\tt Problem}[differentiate, function]} (see calculation on p.\pageref{exp-calc}, specification on p.\pageref{spec-diff} and program on p.\pageref{prog-diff}). The second and third line (b,c) come from {\it{\tt Problem } [on\_interval, goniometric, equation]}: the latter is the postcondition stating that the equation has exactly one solution within the interval given. A single solution is required for this type of problem at this place, and it is the responsibility of the math author to prepare a specification with an appropriate interval. Line (d) is the postcondition of {\tt Subproblem} {\it[on\_interval, goniometric, equation]} on p.\pageref{exp-calc}.

The reader may note that contexts do not follow the scoping rules for program variables (which enclose variables within subprograms): {\it{\tt Problem} ( Reals, [on\_interval, max, argument]} exports the values and the post-conditions of the respective sub-problems (lines (b,c)), {\it{\tt Problem}[differentiate, function]} and {\it{\tt Problem } [on\_interval, goniometric, equation]}. The scoping rules for contexts are a consequence of the common practice to have unique variable names within a calculation.

Although there is (still) no implementation of proofs for postconditions in ISAC{}, there is the conviction that the final context $x_4$ contains sufficient facts for such proofs.

\subsection{Lucas-Interpretation Combines Computation and Deduction}\label{lucas-interp}
A Lucas-Interpreter stepwise executes a CTP-based program (\S\ref{ctp-based-prog}) and creates calculations (Def.\ref{def-calc}) step by step (Def.\ref{def-calc-step}). In order to integrate these concepts, the following definition is close to Def.\ref{def-calc}, Def.\ref{def-calc-step} and Def.\ref{def-result}; actually the only additions are a program and a program state:

\begin{definition}[Lucas-Interpreter]\label{def-exstep}
Given a specification $s$, a calculation $c=\langle X,{\cal P}(F),T,\rightarrow,S\rangle$ and a program $m$, then a labeled terminal transition system ${\cal L}=\langle\Xi,T_m,\rightarrow_m,S_m\rangle$ is called a Lucas-Interpreter iff
\begin{itemize}
\item[(i)] the configuration $\Xi=(\Gamma\times X\times {\cal P}(F))$ contains $\Gamma$ the program states (Def.\ref{def-interp-step}) and $X\times {\cal P}(F)$ the configurations in $c$ (Def.\ref{def-calc})
\item[(ii)] the actions $T_m$ contain (program) tactics in $m$ and $T_m$ is bijectively mapped with $T$
\item[(iii)] the transition relations $\rightarrow^{t_m}$ are steps of interpretation according to Def.\ref{def-interp-step} with $\gamma,x,F\rightarrow^{t_m}\gamma^\prime,x^\prime,F^\prime$ and are ${it initialized\_by}\;s,m$.
\item[(vi)] the terminal configuarions $S_m\subset(\Gamma_m\times X\times {\cal P}(F))$ contain $\Gamma_m$ the terminating states of $m$. 
\end{itemize}
We say $\mathbf{c}\;\textbf{is\_generated\_by}\;\mathbf{(s,m,{\cal L})}$.
\end{definition}
The bijective mapping between $T$ and $T_m$ continues the bijective mapping between {\em external} tactics and {\em internal} tactics from p.\pageref{ext-int-tac}; again further distinction seems not necessary and we write $t$ instead $t_m$ in the sequel.

\medskip
The steps of interpretation in the above definition are triggered by the user requesting \textit{do\_next} according to Def.\ref{def-interp-step}. Instead of \textit{do\_next} the user also could have input a step of his or her own choice; after such input \textit{do\_next} raises the usual issue for debuggers: in which cases can execution (in our case: interpretation) resume, and in which cases it cannot resume due to too invasive operations at the break-point. A step can be given by two classes of input, which might be modified and combined by a dialog component, see \cite{kremp.np:assess}. The first class is characterized by input of an (external) tactic, the second by input of a formula:

\paragraph{Input of a tactic} is done by the learner at steps where the theorem to be applied is the point (e.g. a rewrite rule applied to a huge term too laborious for input), where some sub-term needs to be picked out for the next step, where a sub problem has to be started, or other cases, where the learner prefers to input a tactic instead of a formula (or the dialog component decides according to some rule in the dialog mode). The following definition describes how Lucas-Interpretation handles the input (external) tactic $t_I$:

\begin{definition}[Locatable tactics]\label{def-locat-tac}
Given a specification $s$, a program $m$, a Lucas-Interpreter ${\cal L}=\langle\Xi,T_m,$ 
$\rightarrow_m,S_m\rangle$ at configuration $\chi=(\gamma,x,F)\in\Xi$, a calculation $c$ with $c\;{\it is\_generated\_by}\;(s,m,{\cal L})$ at configuration $x,F$ and finally given an (external) tactic $t_I$ input by the learner, then we have the rule
$$
\displaylines{\hfill
{{  t_I\;{\it applied\_to}\;(x,F)=(x^\prime,F^\prime)  
  \qquad           
    \gamma,x,F
      \rightarrow^*
    \gamma^{\prime},x,F
      \rightarrow^{t_m}
    \gamma^{\prime\prime},x^{\prime},F^{\prime}}
  \qquad
    t_I\approx t_m
\over 
 {\gamma,x,F \rightarrow^{t_m} \gamma^{\prime\prime},x^{\prime},F^{\prime}}}\hfill  
}$$ 
If the rule is applicable (in particular if for $t_I$ an equivalent $t_m$ in the program is found, $t_I\approx t_m$), we say '$\mathbf{t}\; \textbf{is\_locatable\_at}\; \mathbf{\chi}$' iff applicable; otherwise we say $\mathbf{\cal L}\;\textbf{is\_helpless\_at}\;\mathbf{\chi}$.
\end{definition}
The definition's rule has $t_I\;{\it applied\_to}\;(x,F)$ as premise: if this premise is given, we get a step of {\em calculation} leading to $x^{\prime},F^{\prime}$, but if no $t_I\approx t_m$ is found during interpretation, we don't get a $\gamma^{\prime\prime}$, a program state to resume interpretation form --- ${\cal L}$ is helpless.

Note that the transitions $\rightarrow^*\dots\rightarrow^t$ searching for $t_I\approx t_m$ only change the program state $\gamma$ and not the configuration $x,F$ of the calculation; so $x^\prime,F^\prime$ resulting from $t_I$ is presented to the user, before control is handed over. This behavior is consistent with transitions being functions and not relations in both definitions, in calculations (Def.\ref{def-calc}) as well as in Lucas-Interpretation (Def.\ref{def-exstep}). And the strategy implemented by Def.\ref{def-locat-tac} works particularly well with programs like the example on p.\pageref{prog-diff}.

If a $t_m$ with $t_I\approx t_m$ can{\em not} be found, the step of calculation can still be done (according to \\$t_I\;{\it applied\_to}\;(x,F)$) --- but no appropriate tactic could be found in the program, thus the Lucas-Interpreter cannot determine a next step and ``is helpless''. Since this case easily can happen (for instance, if a ``creative'' rewrite rule is applied which cannot be found in the program), there is a stronger feature of Lucas-Interpretation for input of formulas.

\paragraph{Input of a formula} concerns the second class of input for a learner constructing a calculation. Since user interfaces are expected to support copy-and-paste, input need not be laborious even for huge formulas. The following definition describes how Lucas-Interpretation handles an input formula $f_I$:
\begin{definition}[Derivable formula]\label{def-deriv-form}
Given a specification $s$, a program $m$, a Lucas-Interpreter ${\cal L}=\langle\Xi,T_m,$ 
$\rightarrow_m,S_m\rangle$, a calculation $c$ with $c\;{\it is\_generated\_by}\;(s,m,{\cal L})$ at configuration $x,F$ and finally given a formula $f_I$ input by the learner, then we have the rule
$$
\displaylines{\hfill
{{\gamma,x,F \rightarrow^* \gamma^\prime,x^\prime,F^\prime  \qquad  F\vdash_{x^\prime} f_I}
\over 
 {\gamma,x,F \rightarrow^{t^*} \gamma^\prime,x^\prime,F_I}}\hfill  
}$$
If the rule is applicable we say '$\mathbf{f_I\; \mathbf{is\_derived\_from}\; \mathbf\chi}$'; otherwise we say '$\mathbf{f_I}\;\mathbf{is\_not\_derivable\_from}\; \mathbf\chi$'.
\end{definition}
In this case the interpreter executes the next tactics found in program $m$ until a context $x^\prime$ is generated, which can justify the input $f_I$, i.e. $c\vdash_{x^\prime} f_I$. This justification might involve algebraic simplification as mentioned in \S\ref{deduct}. If successful, the step is presented to the user, before control is handed over; $t^*$ is a tactic called ``ad-hoc derivation''; it usually comprises a sequence of tactics.

If not successful, i.e. the interpreter executes program $m$ until termination and no step is found with $c\vdash_{x^\prime} f_I$, then $f_I$ is ``not derivable''. This judgment can be costly in resources, since it relies on an (internal) computation of the program, including all the subprograms, until termination.

\section{Related Work and Open Questions}\label{open-related}

\subsection{Proofs, Structured Derivations and Calculations}\label{logic-foundation}
As mentioned above, educational systems shall be ``systems which do not make mistakes'', so well-designed logical foundations are indispensable.

\paragraph{Proof languages} like Isar \cite{mw:isar07} are natural to compare with the approach taken in this paper, since the underlying prototype is based on the CTP Isabelle \cite{Nipkow-Paulson-Wenzel:2002}. Unlike the proof languages of Coq \cite{Huet_all:94}, Ltac \footnote{http://coq.inria.fr/stdlib/Coq.Program.Tactics.html} and SSReflect \cite{DBLP:conf/ascm/Gonthier07},
Isar is self-contained and does not refer to a proof state explicitly. 

The requirement to be ``as close to what is written to traditional blackboards as possible'' inhibited the above calculations to follow the principle of being self-contained. Rather, calculations rely on {\em internal} contexts employed in all definitions. This is the price for high usability for the target group. The price is considered not too high, since contexts consist of formulas in predicate logic and thus can easily be presented to the user, as soon as the user requests for them (and then they can be filtered by a dialog component).

Isar specifically supports calculational proof \cite{makar:calc01} and integrates it well with other elements of the proof language. Nevertheless, the formalisms were considered too far from ``what is written on traditional blackboards'' for our target group.

\paragraph{Structured derivations} (SD) \cite{Back-SD09} continue work on calculational proof \cite{bk:dij90,gries:calc-proof}, provide foundations on natural deduction, are self-contained such that they need not refer to a state separate from the SD and also are ``close to what is written on blackboards''.

So, SD meet essential requirements for mechanized calculation and call for justification of any other choice of format. Major parts of the example calculation on p.\pageref{exp-calc} exactly model SD, for instance lines {\rm \#08..\#13}. The sub-{\tt Problem}s strengthen the rigor by requiring explicit formal specification (which might be hidden from the learner and done automatically behind the scenes). And the rigor of SD is weakened by allowing to omit the tactics (at the right margin on p.\pageref{exp-calc}). The weakened rigor allows to address a wider range of users, who need no prior introduction to the system --- ``next step guidance'' provides perfect ad-hoc introduction. Another point is, that SD concerning sub-terms of formulas involve ``window inferencing'' \cite{luth.ea:tas:2000}, which is already more complicated.

For these reasons a format more general than SD has been implemented in ISAC.

\subsection{``Correctness'' in Lucas-Interpretation}\label{correctness}
This subsection doesn't address related work, only open questions on ``correctness''. Correctness of calculations is indispensable for an educational ``system which never makes mistakes'' \cite{cezary-phd}. Final correctness of a calculation is given by Def.\ref{def-result}: the solution meets the postcondition. The steps (tactic/formula) of calculation constructing such a solution are either input by the user or determined by a program under Lucas-Interpretation. An input tactic is checked if it \textit{is\_applicable\_in} the calculation (Def.\ref{def-calc-step}) -- this implies ``correctness''. An input formula is proven derivable (Def.\ref{def-deriv-form}) or not, thus it is ``correct'' or not. Both notions of ``correctness'', however, cannot exclude steps which are ``misleading'' to somewhere and not to a solution. But if the user relies on the system and requests \textit{do\_next} (Def.\ref{def-interp-step}), the system shall guarantee to come to a solution finally: 

\begin{definition}[Correctness under Lucas-Interpretation]\label{correct-LI}
Given a specification $s$, a program $m$, a Lucas-Interpreter ${\cal L}=\langle\Xi,T_m,\rightarrow_m,S_m\rangle$, a calculation $c$ with $c\;{\it is\_generated\_by}\;(s,m,{\cal L})$ and a sequence of steps of interpretation triggered by {\it do\_next} $\chi_0\rightarrow^{\it do\_next}\dots\rightarrow^{\it do\_next}\chi_n=(\gamma_n,x_n,F_n)$. Then $c$ is correct under $s,m,{\cal L}$ iff
$$\chi_n\in S_m\Rightarrow c\;{\it is\_completed\_with}\;(x_n,F_n)$$
\end{definition}
This definition says: if Lucas-Interpretation terminates, the generated calculation provides a solution $(x_n,F_n)$. So we only have a definition of correctness, but we want to have a proof! The definition describes the general proof obligation for the programmer of $m$ and $s$. In order to actually verify $m$ with respect to $s$ a rigorous formal semantics for each tactic $t$ will be required: such a semantics just would have to add specific properties of $t$ to Def.\ref{def-calc-step} (which then would carry over to Def.\ref{def-interp-step}). This theoretical work seems most promising if combined with RTD of a development environment which integrates specific verification tools: a CTP-based IDE for a CTP-based programming language.

ISAC already shows the advantageous consequences of the above definition: the learner, if got stuck after some trials, always can backtrack to a step where reaching a solution is guaranteed --- this feature allows interactive learning known from chess programs: when reaching a disadvantageous chess configuration just backtrack to another configuration and start a new trial (a novel kind of interaction for EMAs). Experience with the prototype shows, that learners get used to find out the position where to back-track very quickly.


\medskip
The open questions above on correctness have already raised vague ideas, that the components involved in Lucas-Interpretation are related by refinements \cite{Back1998}. The components are specification, calculation, problem-class and program (where problem-class has not yet been discussed):

{\small
\begin{center}
\begin{tabular}{|l l l l|} \hline
&&&$\;$\\      
\hspace{0.3cm}(1.)
     &{\it specification}
          & $\sqsubseteq$
               &\hspace{0.2cm}{\it calculation}\hspace{0.2cm}$\;$\\
\hspace{0.3cm}$\longrightarrow$&&&$\;$\\
\hspace{0.3cm}&$\sqsupseteq$&&\hspace{0.2cm}$\sqsupseteq$\\
&&&$\;$\\
\hspace{0.3cm}(2.)\hspace{0.1cm}
     &{\it problemclass}
          & $\sqsubseteq$
               &\hspace{0.2cm}{\it program}\\
\hspace{0.3cm}$\longrightarrow$&&&$\;$\\
&$\uparrow\;\;$(3.)&&\hspace{0.2cm}$\uparrow\;\;$(4.)\\
\hline
\end{tabular}
\end{center}
}

A look at the above refinement relations ($\sqsubseteq$) gives the following big picture:
\begin{enumerate}
\item${\it specification}\sqsubseteq{\it calculation}$ seems clear to a mathematician; however, the underlying definitions (Def.\ref{def-calc-step}, Def.\ref{def-calc} and Def.\ref{def-result}) might be not strong enough for describing this refinement.
\item\label{pblclass}${\it problemclass}\sqsubseteq{\it program}$ involves a notion not introduced here because not yet clarified: ``problemclass'' would be some ``specification-pattern'' which allows to mechanically generate a specification of the program from the input 'props': see example specifications no.1 and no.2 and imagine all the other examples belonging to the problemclass ``Extremwert Aufgaben'' (which is claimed to be solved by one and the same program, the example program) --- how would a specification (-pattern), as general as the program, would look like?
\item${\it problemclass}\sqsubseteq{\it specification}$ is expected to help clarifying the question from the previous Pt.\ref{pblclass}.
\item${\it program}\sqsubseteq{\it calculation}$ is the crucial point addressed by Def.\ref{correct-LI}.
\end{enumerate}

\subsection{On CTP-based Programming Languages}\label{ctp-based-lang}
%
%

Lucas-Interpretation takes a certain kind of ``CTP-based programming language'' as a prerequisite, which only exists in ISAC{} under consideration. However, \cite{plmms10} argue for a value of their own for such languages and identify interest for CTP-based languages: For CAS-based programming languages great demand appears in successful application to software for engineering and science, most of which use CAS-based languages like \cite{prog-maple06,progr-mathematica} as an efficient base for development --- while lacking reliability of CAS is widely recognized, see for instance \cite{harr:book}; so one may expect developments meeting the demand indicated by the success of CAS-based languages with more reliable CTP-based languages.

Significant preparatory work \cite{caspartial,casproto} for CAS-like functionality in a CTP framework has already been done. In the course of eventual development of CTP-based languages several further issues left open by CAS might be tackled: Reliable handling of partiality conditions has been mentioned in \S\ref{cas-like-funct}. Multivaluedness is another issue \cite{seeingroots}. Very appealing for educational purposes would be to calculate exact approximation given by floating-point numbers (see \cite{russellphd} for Coq) throughout all algebraic operations involved in a calculation; already \cite{harr:book} has laid the grounds for tackling this issue.

\medskip
Using CTP-based programs not only for implementation of applications like the running example (p.\pageref{prog-max}), but also for the CAS-functions themselves will lead to ``systems that explain themselves'' at an extreme level: For instance, cancellation of multivariate polynomials like
$$\frac{x^2-y^2}{x^2-x\cdot y}=
\frac{(x+y)\cdot(x-y)}{x\cdot(x-y)}=
\frac{x+y}{x}\;\;\;\;{\it assuming}\;x\not=0\land x-y\not=0 $$
is a frequent task already in early years of math education; usually it is explained ``from right to left''. But a smart student might ask, how a machine accomplishes the task from left to right (by factorization, by Hensel lifting or the like). If factorization or Hensel lifting is programmed in a CTP-based language and interpreted by a Lucas-Interpreter, then such a student gets stepwise user guidance through the comprehensive algorithms on request:

\medskip
{\em With a CTP-based language and Lucas-Interpretation ``implementation of tutoring becomes a side-effect of programming''}, see Fig.\ref{fig.lucin-sideeffect} on p.\pageref{fig.lucin-sideeffect}.

With that in mind interesting questions arise when re-implementing CAS-functions: CAS-algorithms at the state-of-the-art are high-brow in general (not only in the above case, also in integration, solving equational systems etc.). In order to foster students' curiosity and do not bother them upon questions of ``how is that done'', the possibility to select between high-brow algorithms and elementary algorithms (the latter probably not able to solve a certain problem) is already implemented in ISAC{}.

\section{Summary and Conclusions}\label{summ-concl}
This paper presented an innovation in base technology for educational math assistants which is mainly motivated by pedagogical requirements. So we finish with a summary of technology and a conclusion concerning pedagogy.

\paragraph{A Summary of Lucas-Interpretation} states a novel base technology for automatic generation of user guidance in stepwise solving problems in engineering and in applied science: Given a program written in a novel kind of CTP-based programming language which solves a problem, Lucas-Interpretation generates steps (Def.\ref{def-calc-step}, Def.\ref{def-interp-step}) of a calculation (Def.\ref{def-calc}) as a side-effect (Fig.\ref{fig.lucin-sideeffect} on p.\pageref{fig.lucin-sideeffect}) of the (functional) program. During stepwise interpretation a logical context is built up in order to provide ATP tools with the facts required for proving derivability of user input. At each tactic in the program control is handed over to the user (Def.\ref{def-exstep}). The user is free to investigate the underlying knowledge, to freely input a step independently (checked by ATP immediately) or to ask the system to propose a next step (Def.\ref{def-locat-tac}, Def.\ref{def-deriv-form})

According to the novelty of the design essential questions have been left open and are addressed in a separate section: In particular, the main theorems of Lucas-Interpretation are not proved due to the lack of a formalized operational semantics: The solution of a calculation is correct if the user does not interfere by other input than requesting the next step (Def.\ref{correct-LI}) and interpretation can guarantee to resume after user input provided certain constraints on the input.

Although still a prototype, the Lucas-Interpreter has already established a stable interface to a dialog component, ready for experts in learning theory to start with detailed design of interaction in stepwise problem solving --- this task is rigorously separated from extending the math topics of the problems: each problem just requires to program the algorithm solving the problem, interaction is handled by Lucas-Interpretation.

\paragraph{Conclusions for pedagogy} expect impact on math education: CTP-technology allows to relate formal facts and activities in stepwise problem solving {\em mechanically within} the system. This is a considerable advantage over CAS or DGS (Dynamic Geometry Systems), where specific steps (integration, equation solving, etc.) are accomplished by the systems, well, but the whole process of stepwise problem solving, the relation between input (precondition), output (postcondition) and intermediate steps are left to the programmer media designer, who is overwhelmed by hard-coding feedback for the variants of steps.

On the other hand, CTP-based software models all mechanical parts of problem solving and thus allows to mechanically generate feedback to the learner: ATP ensures both, the most reliable {\em and} the most general checks of user input. This extends the scope of EMAs far beyond CAS or DGS. {\em So upcoming CTP-based educational math assistants will be novel and powerful tools for patient and reliable feedback in exercising and assessing stepwise problem solving}. 

\medskip
If, in addition, educational math assistants want to go in line with renewed science education \cite{rocard07-short} and decisively support inquiry-based learning and independent learning, then {\em``next step guidance''} is required: students can explore {\em new} topics, tackle motivating problems and interactively try out solutions passing new stuff by help of the system, students can try variants in calculations and ask for the next step if got stuck --- the latter is {\em the unique feature Lucas-Interpretation contributes as a base technology for educational math assistants}.

\section*{Acknowledgments}
The author thanks Peter Lucas for granting the approval to connect his name with the ideas presented in this paper, ideas which have not yet gained success comparable with his other research and developments.

Thanks also to Franz Wotawa for his long-lasting support in developing ISAC until usability in educational practice and academic discussion of the underlying ideas have been established.

This paper could not have been written without the help of two colleagues: Bernhard Aichernig, Peter Lucas' former student and his assistant, gave the hints which formalisms to use; last but not least Wolfgang Schreiner, RISC Linz, greatly helped to appropriately apply these formalisms.

\bibliographystyle{eptcs}

\end{document}